\documentclass[letter,bibyear]{aa} 

\usepackage{epsfig}
\usepackage{amsmath}
\usepackage{subfigure}
\usepackage{natbib}
\usepackage{multirow}
\usepackage{color}
\usepackage{longtable}
\usepackage{graphicx}
\usepackage{epstopdf}
\usepackage{booktabs}
\usepackage{txfonts}

\newcommand{\jmst}{J.~Mol.~Struct.}

\newcommand{\kms}{km s$^{-1}$}

\begin{document}

\title{Detection of vibrationally excited C$_6$H  in the cold prestellar core TMC-1 with the QUIJOTE line
survey\thanks{Based on 
observations carried out
with the Yebes 40m telescope (projects 19A003,
20A014, 20D023, 21A011, 21D005, and 23A024). The 40m
radio telescope at Yebes Observatory is operated by the Spanish Geographic Institute
(IGN, Ministerio de Transportes, Movilidad y Agenda Urbana).}}

\author{
J.~Cernicharo\inst{1},
R.~Fuentetaja\inst{1},
M.~Ag\'undez\inst{1},
C.~Cabezas\inst{1},
B.~Tercero\inst{2,3},
N.~Marcelino\inst{2,3}, and
P.~de~Vicente\inst{3}
}

\institute{Dept. de Astrof\'isica Molecular, Instituto de F\'isica Fundamental (IFF-CSIC),
C/ Serrano 121, 28006 Madrid, Spain. \newline \email jose.cernicharo@csic.es,r.fuentetaja@csic.es
\and Observatorio Astron\'omico Nacional (OAN, IGN), C/ Alfonso XII, 3, 28014, Madrid, Spain.
\and Centro de Desarrollos Tecnol\'ogicos, Observatorio de Yebes (IGN), 19141 Yebes, Guadalajara, Spain.
}

\date{Received 05/11/2023; accepted 17/11/2023}

\abstract{
In this work, we present the detection of twelve doublets with quantum numbers of $N$=12-11 to $N$=17-16 of 
the $\nu_{11}(\mu^2\Sigma$) vibrationally excited state of C$_6$H  towards TMC-1. This marks the first time that an excited vibrational state of a molecule
has been detected in a cold starless core. The data are part of the QUIJOTE line survey gathered with the Yebes 40m radio telescope. 
The line intensities have been aptly reproduced with a rotational temperature of 6.2$\pm$0.4\,K 
and a column density of (1.2$\pm$0.2)$\times$10$^{11}$ cm$^{-2}$. We also analysed the ground state transitions of 
C$_6$H, detecting fourteen lines with quantum numbers of $J$ = 23/2-21/2 to $J$ = 35/2 for each of the two 
$^2\Pi_{3/2}$ and $^2\Pi_{1/2}$ ladders. It is not possible to model the intensities of all the transitions of the ground state 
simultaneously using a single column density. We considered the two ladders
as two different species and found that the rotational temperature is the same for both ladders, 
T$_{rot}(^2\Pi_{3/2})$=T$_{rot}(^2\Pi_{1/2})$=6.2$\pm$0.2, achieving a result that is comparable to that of the $\nu_{11}(^2\mu\Sigma)$ state. The derived column densities are
N($^2\Pi_{3/2}$)=(6.2$\pm$0.3)$\times$10$^{12}$ cm$^{-2}$ and N($^2\Pi_{1/2}$)=(8.0$\pm$0.4)$\times$10$^{10}$ cm$^{-2}$.
The fraction of C$_6$H molecules in its 
$^2\Pi_{3/2}$, $^2\Pi_{1/2}$, and $\nu_{11}(\mu^2\Sigma$) states  is 96.8\,\%, 1.3\,\%, and 1.9\,\%, respectively. 
Finally, we report that this vibrational mode has also been detected towards the cold cores Lupus-1A and L1495B, as well as the low-mass
star forming cores L1527 and L483, with
fractions of C$_6$H molecules in this mode  of 3.8\,\%, 4.1\,\%, 14.8\,\%, and 6\,\%, respectively.
}

\titlerunning{Vibrationally C$_6$H in TMC-1}
\authorrunning{Cernicharo et al.}

\maketitle

\section{Introduction}
The first detection of C$_6$H in space has been achieved towards TMC-1 \citep{Suzuki1986,Cernicharo1987a} and in the 
direction of the carbon rich star
IRC+10216 \citep{Guelin1987}. The identification was based on ab initio calculations of the
structure and electronic state of the molecule, which indicate that the ground state could be 
a $^2\Pi$ state \citep{Murakami1988}, contrary to CCH and C$_4$H, which have a $^2\Sigma^+$ ground state. The detection in space of two
series of harmonically lines with half-integer quantum numbers 
\citep[lines from the $^2\Pi_{3/2}$ and
the $^2\Pi_{1/2}$ ladders,][]{Suzuki1986,Guelin1987,Cernicharo1987a} clearly supported the identification. Nevertheless, the
assignment of the observed lines to this species was confirmed by the generation of the
radical and the observation of its rotational spectrum in the laboratory \citep{Pearson1988}. The spin orbit constant,
$A_{SO}$, is $-$15.04 cm$^{-1}$ \citep{Linnartz1999}. Hence, the energy of the $^2\Pi_{1/2}$ ladder is $\sim$22\,K above the
$^2\Pi_{3/2}$ one.

C$_6$H has six stretching modes, $\nu_1-\nu_6$, and five bending modes, $\nu_7-\nu_{11}$ \citep{Brown1999}. 
The $\nu_{11}$ mode is the bending with the lowest energy, which has been estimated to be between 120 and 134 cm$^{-1}$ \citep{Brown1999,Cao2001}. 
This bending mode produces a $^2\Delta$ and two $^2\Sigma$ vibronic modes. One
of the $^2\Sigma$ states, labelled as $\mu^2\Sigma$ by \citet{Zhao2011}, and the $^2\Delta$ one have been identified in IRC+10216 
during a search for C$_5$N$^-$
\citep{Cernicharo2008}. The assignment was subsequently  confirmed by extensive laboratory spectroscopy of C$_6$H in its $\nu_{11}$ vibrationally
excited state \citep{Gottlieb2010}. 

The $\nu_{11}$ vibrational mode is affected by a strong Renner-Teller interaction
resulting from the coupling of the degenerate bending vibration with the orbital angular momentum of
an unpaired electron \citep{Zhao2011}. This coupling produces a considerably lowering of
the energy of the $\mu^2\Sigma$ vibronic mode. This energy has been estimated to be 20$\pm$10\,K \citep{Gottlieb2010}. In a detailed study involving
the electronic transition B$^2\Pi$-X$^2\Pi$, this energy has been determined to be 11.0$\pm$0.8\,cm$^{-1}$ \citep{Zhao2011}, namely, around 15.8\,K.
This value is below the energy the $^2\Pi_{1/2}$ ladder of the ground electronic state. Hence, this
vibronic mode of the $\nu_{11}$ vibrational state could be detected in sensitive line surveys of cold sources.

TMC-1 is a cold starless core located in Taurus at a distance of 140 pc \citep{Cernicharo1987b}. Its kinetic temperature is 10\,K and,
so far, all observed lines correspond to rotational transitions of a large variety of molecular species in their ground vibrational state.
Using the sensitive QUIJOTE\footnote{\textbf{Q}-band \textbf{U}ltrasensitive \textbf{I}nspection \textbf{J}ourney to the \textbf{O}bscure 
\textbf{T}MC-1 \textbf{E}nvironment} line survey \citep{Cernicharo2021a}, we report the detection in TMC-1 of 12 lines of the $\nu_{11}(\mu^2\Sigma)$ vibrational mode of C$_6$H, and of 28 lines of
the $^2\Pi_{3/2}$ and  $^2\Pi_{1/2}$ ladders of its ground electronic state. 
It is the first time that emission from a vibrationally excited state has been detected in a cold astrophysical environment.

\section{Observations}
The observational data used in this work are part of QUIJOTE \citep{Cernicharo2021a}, 
a spectral line survey of TMC-1 in the Q-band carried out with the Yebes 40m telescope at 
the position $\alpha_{J2000}=4^{\rm h} 41^{\rm  m} 41.9^{\rm s}$ and $\delta_{J2000}=
+25^\circ 41' 27.0''$, corresponding to the cyanopolyyne peak (CP) in TMC-1. The receiver 
was built within the Nanocosmos project\footnote{\texttt{https://nanocosmos.iff.csic.es/}} 
and consists of two cold high-electron mobility transistor amplifiers covering the 
31.0-50.3 GHz band with horizontal and vertical polarisations. Receiver temperatures 
achieved in the 2019 and 2020 runs vary from 22 K at 32 GHz to 42 K at 50 GHz. Some 
power adaptation in the down-conversion chains have reduced the receiver temperatures 
over the course of 2021 to 16\,K at 32 GHz and 30\,K at 50 GHz. The backends are 
$2\times8\times2.5$ GHz fast Fourier transform 
(FFT) spectrometers with a spectral resolution of 38 kHz, providing the whole coverage 
of the Q-band in both polarisations.  A more detailed description of the system 
is given by \citet{Tercero2021}. 

The data of the QUIJOTE line survey presented here were gathered in several 
observing runs between November 2019 and July 2023.  
All observations were performed using frequency-switching observing mode with 
a frequency throw of 8 and 10 MHz. The total observing time on the source 
for data taken with frequency throws of 8 MHz and 10 MHz is 465 and 737 hours, 
respectively. Hence, the total observing time on source is 1202 hours. The 
measured sensitivity varies between 0.08 mK at 32 GHz and 0.2 mK at 49.5 GHz.  
The sensitivity of QUIJOTE is around 50 times better than that of previous 
line surveys in the Q band of TMC-1 \citep{Kaifu2004}. For each frequency 
throw, different local oscillator frequencies were used in order to remove 
possible side band effects in the down conversion chain. A detailed description 
of the QUIJOTE line survey is provided in \citet{Cernicharo2021a}.
The data analysis procedure has been described in \citet{Cernicharo2022}.
The main beam efficiency measured during our observations in 2022 
varies from 0.66 at 32.4 GHz to 0.50 at 48.4 GHz \citep{Tercero2021} and can be given across the Q-Band by
$B_{\rm eff}$=0.797 exp[$-$($\nu$(GHz)/71.1)$^2$]. The
forward telescope efficiency is 0.97.
The telescope beam size at half power intensity is 54.4$''$ at 32.4 GHz and 36.4$''$ 
at 48.4 GHz. 

The intensity scale 
utilised in this study is the antenna temperature ($T_A^*$). 
Calibration was performed using two absorbers at 
different temperatures and the atmospheric transmission model ATM \citep{Cernicharo1985, 
Pardo2001}. The absolute calibration uncertainty is 10$\%$. However, the relative
calibration between lines within the QUIJOTE survey is probably better because all
the spectral features have common pointing and calibration errors.
The data were analysed with the GILDAS package\footnote{\texttt{http://www.iram.fr/IRAMFR/GILDAS}}.

\begin{figure}[h] 
\centering
\includegraphics[width=0.485\textwidth]{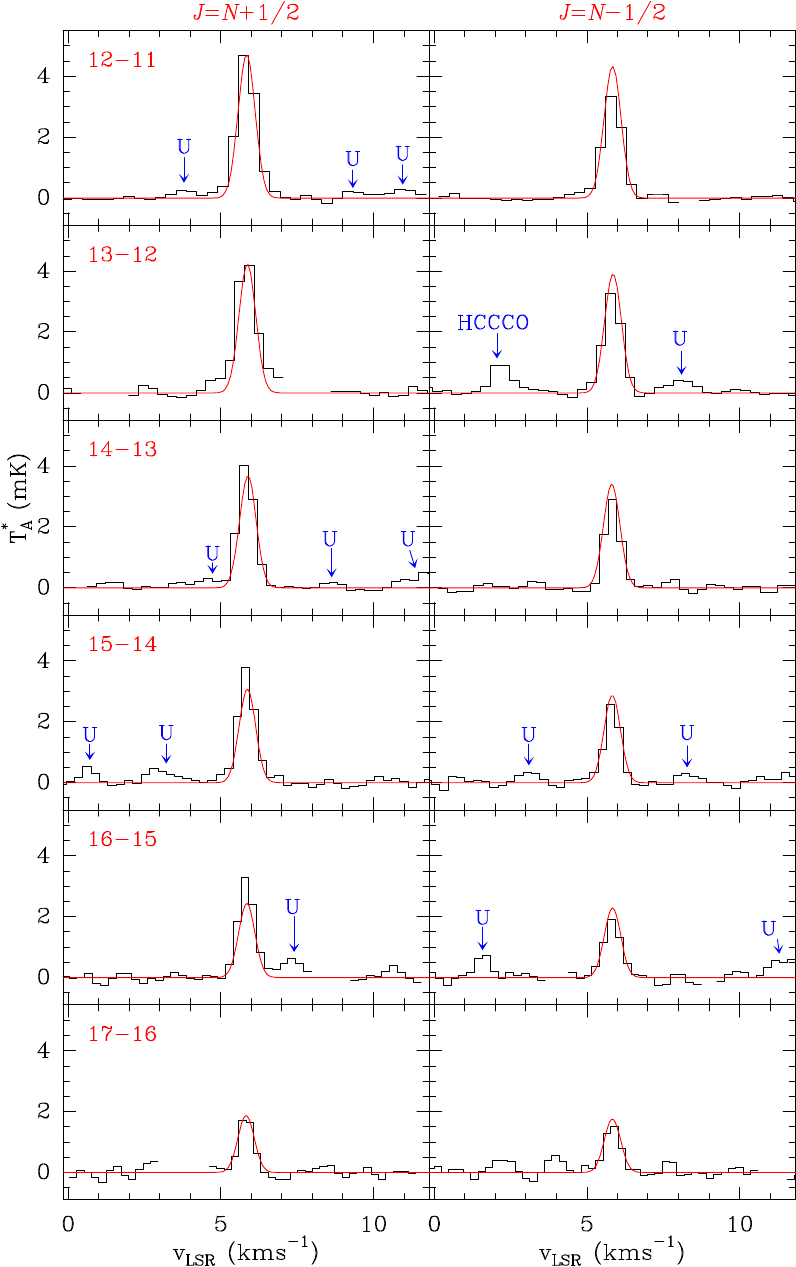}
\caption{
Selected transitions for the excited state of C$_6$H in TMC-1. Left and right panels correspond to the
$J=N+1/2$ and $J=N-1/2$ components of each transition, respectively.
The abscissa corresponds to the local standard of rest (LSR) velocity. The derived line parameters
are given in Table \ref{line_parameters}. The ordinate is the antenna temperature, corrected for 
atmospheric and telescope losses, in mK.
Blanked channels correspond to negative features 
produced when folding the frequency-switched data.
The quantum numbers of each transition are indicated
in the corresponding panel. The red line shows the computed synthetic spectra for this species for 
T$_{rot}$ = 6.2 K and a column density of 1.2$\times$10$^{11}$ cm$^{-2}$.  
} 
\label{C6H_v11}
\end{figure}

\begin{figure*}[h] 
\centering
\includegraphics[width=0.49\textwidth]{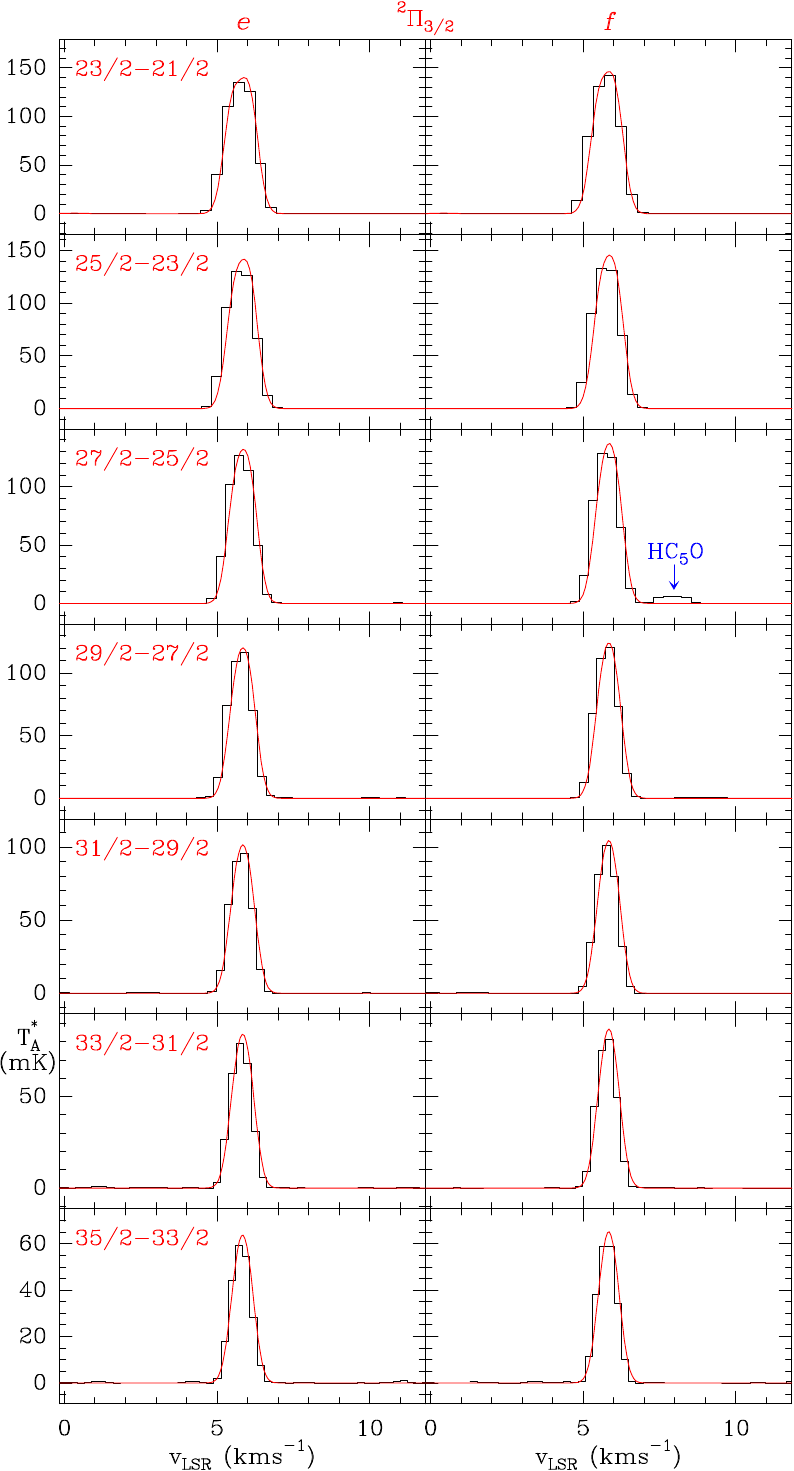}
\includegraphics[width=0.49\textwidth]{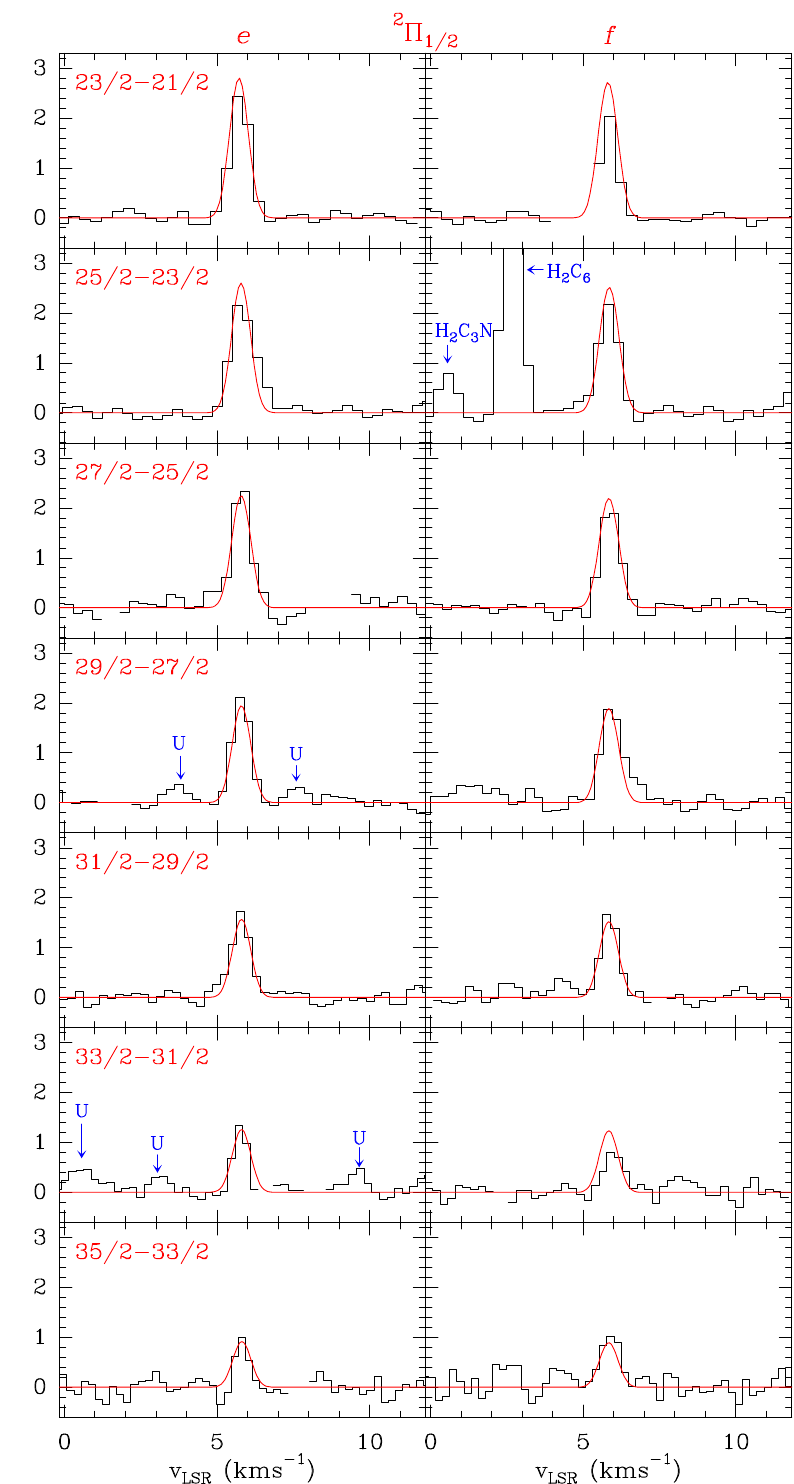}
\caption{Observed transitions in TMC-1 of the $^2\Pi_{3/2}$ (two left columns) and
$^2\Pi_{1/2}$ (two right columns) states of C$_6$H. 
The labels 
$e$ and $f$ correspond to the Lambda-type doubling components of each transition.
The abscissa corresponds to the LSR velocity. The derived line parameters
are given in Table \ref{line_parameters}. The ordinate is the antenna temperature, corrected for 
atmospheric and telescope losses, in mK.
Blanked channels correspond to negative features 
produced when folding the frequency-switched data.
The quantum numbers of each transition are indicated
in the corresponding panel.
The red line shows the computed synthetic spectra for this species (see Sect. \ref{sec:results}).
}
\label{C6H}
\end{figure*}

\section{Results}\label{sec:results}

The line identification in this work was performed using the MADEX code \citep{Cernicharo2012} and the CDMS catalogue \citep{Muller2005}.
The spectroscopy of C$_6$H in the ground electronic state is directly adopted from the CDMS catalogue. 
C$_6$H has been previously observed in its ground electronic state in TMC-1 \citep{Suzuki1986,Cernicharo1987a}. 
The $\nu_{11}(\mu^2\Sigma)$ state
is highly perturbed and high distortion constants are needed to reproduce the laboratory frequencies \citep{Gottlieb2010}.
We fit all laboratory frequencies with distortion constants $D, H, L,$ and $M$ for the rotation, as well as $D$ and $H$ for the
fine structure. The results have been implemented into the MADEX code to predict the frequencies of its rotational and fine structure
transitions (see Appendix \ref{rota_v11}).
The adopted dipole moment of the molecule is 5.54\,D \citep{Woon1995}. We have asumed that it is the same for the
$\nu_{11}(\mu^2\Sigma)$ vibrationally excited state.
However, the dipole moment for the rovibrational transitions $\nu_{11}(\mu^2\Sigma)$\,$\rightarrow$\,$X^2\Pi$ is unknown. Moreover,
the corresponding frequencies would lie in the millimeter and submillimeter frequency domains.

\begin{figure*}
\centering
\includegraphics[width=0.95\textwidth]{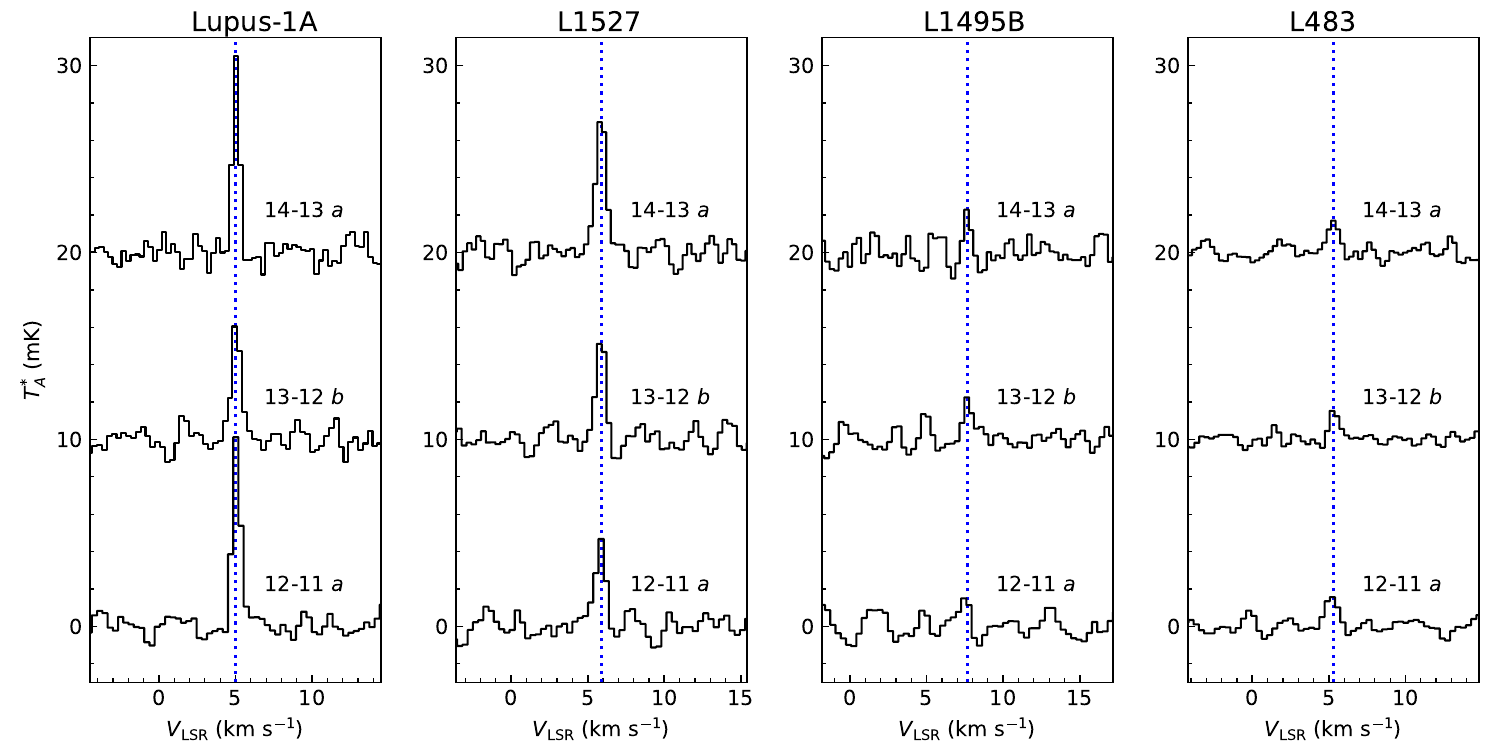} \label{fig:lines_clouds}
\caption{Selected transitions of C$_6$H $\nu_{11}$($\mu^2\Sigma$) observed in Lupus-1A, L1527, L1495B, and L483 with 
the Yebes\,40m telescope. Abcisa is the LSR\ velocity in km\,s$^{-1}$. The ordinate
is the antenna temperature corrected for telescope and atmospheric losses. Line parameters for all observed
lines are given in Table \ref{table:lines_clouds}.} 
\end{figure*}

We detected a total of 40 lines of C$_6$H within the QUIJOTE line survey. Their intensities range from 148 mK to $\sim$1 mK. 
The derived line parameters are given in Table \ref{line_parameters}.
Among these lines, we detected 12 transitions of the $\nu_{11}(\mu^2\Sigma)$ vibrationally excited state. They involve quantum numbers from 
$N$=12-11 to $N$=17-16. For each rotational transition, two fine structure components were detected ($J=N+1/2$ and $J=N-1/2$; denoted as
$a$ and $b$ in Tables \ref{line_parameters} and \ref{table:lines_clouds}). The
lines are shown in Fig. \ref{C6H_v11}. The hyperfine splitting of the lines of
$\nu_{11}(\mu^2\Sigma)$ state is  smaller than 
that of the lines in the ground state and does not affect the line widths.
Fourteen of the detected lines correspond to the $^2\Pi_{3/2}$ ladder of the ground electronic state. They are the strongest transitions of C$_6$H and their 
quantum numbers range from $J$ = 23/2-21/2 to $J$ = 35/2-33/2. 
For the $^2\Pi_{1/2}$ ladder, we also detected 14 lines involving the same quantum numbers. They are are the weakest lines of C$_6$H, that is, even weaker than those
of the $\nu_{11}(^2\Sigma)$  state. The two ladders of the
ground electronic state show two Lambda-type doubling components for each rotational transition
(denoted as $e$ and $f$ in Table \ref{line_parameters}
and in Fig. \ref{C6H}).
The observed lines of the ground state are shown in Fig. \ref{C6H}.
These lines are slightly broadened
due to the hyperfine structure of each line ($\sim$60 and 40 kHz for the $^2\Pi_{3/2}$ and $^2\Pi_{1/2}$ ladders, respectively).
All the 
observed frequencies of the ground state are within 5 kHz of the predicted center frequencies for each transition. 
For the
$\nu_{11}(\mu^2\Sigma)$ mode we have merged the observed frequencies with
the laboratory data of \citet{Gottlieb2010} to get an improved set of rotational, fine structure, and  distortion constants
(see Appendix \ref{rota_v11}).

To estimate the column densities and rotational temperatures, we assumed that the source has a diameter of 80$''$
and that it has an uniform brightness distribution for all observed lines \citep[][Fuentetaja et al. 2023, in prep.]{Fosse2001}. The adopted
intrinsic linewidth is 0.6 km\,s$^{-1}$. The $\nu_{11}(\mu^2\Sigma)$ mode is radiatively connected to the $^2\Pi_{3/2}$ and
$^2\Pi_{1/2}$ ladders of the ground state. All the levels should be treated simultaneously to solve the statistical equilibrium
equations. 

Although collisional rates are available for C$_6$H in its ground electronic state \citep{Walker2018}, no such rates
are available for the $\nu_{11}(\mu^2\Sigma)$ mode. 
The energy of the levels involved in the transitions observed for this state, without adding the energy difference between it 
and the ground state,
varies between 10.4 ($N_u$=12) and 20.5\,K ($N_u$=17). Consequently, $T_{rot}$ should be well constrained by the data.
To obtain the column density and rotational temperature, we used a model line fitting procedure,
adopting a LTE approach (all transitions with identical rotational temperature) in which both parameters are optimised  
(see e.g. \citealt{Cernicharo2021b}). We obtain $T_{rot}$=6.2$\pm$0.4\,K and
$N(\nu_{11}(\mu^2\Sigma)$)=(1.2$\pm$0.2)$\times$10$^{11}$ cm$^{-2}$. Using a standard rotational diagram, we obtained similar results. The
column density for the $\nu_{11}(\mu^2\Sigma)$ state is similar to that of C$_6$H$^-$ \citep{McCarthy2006}
and to that of many other molecules of the C$_n$H$_m$, C$_n$H, C$_n$H$^-$, and C$_n$H$^+$ families previously  analysed with QUIJOTE \citep[Table A.1  of][]{Cernicharo2022}.

For the ground electronic state, we tried to use the collisional rates of C$_6$H, with He calculated by \citet{Walker2018}
and including the two ladders. However,
no satisfactory solutions have been found (see discussion on this point in \citealt{Agundez2023}). To roughly reproduce the intensities of the $^2\Pi_{3/2}$ ladder,
we have to use a value for $n$(H$_2$) that is uncommon for TMC-1, $\sim$10$^6$ cm$^{-3}$, without having still a reasonable fit to the lines of 
the $^2\Pi_{1/2}$ ladder. Rotational temperatures for these high densities are around 6\,K and the column density needed to reproduce
the $^2\Pi_{3/2}$ ladder is $\sim$5$\times$10$^{12}$ cm$^{-2}$. However, the predicted lines of the $^2\Pi_{1/2}$ ladder are
a factor of two stronger than what is observed.  Using the line fitting procedure, with the assumption of a common
rotational temperature for all levels, we found the same problem if both ladders are treated simultaneously. In both cases, LVG or LTE,
the column density that fits the $^2\Pi_{3/2}$ ladder has to be divided by a factor of two to reproduce the intensities of the
$^2\Pi_{1/2}$ one. This indicates
that the rotational and interladder temperatures are different, and/or that the $^2\Pi_{1/2}$ ladder is sharing a fraction of
its population with the $\nu_{11}(\mu^2\Sigma)$ vibrational mode.

To obtain reasonable values for the column densities in the two ladders of the ground vibrational state, we considered
each ladder separately. We derived $T_{rot}$($^2\Pi_{3/2}$)\,=\,6.2\,$\pm$\,0.2 K, $T_{rot}$($^2\Pi_{1/2}$)\,=\,6.2\,$\pm$\,0.4 K,
$N$($^2\Pi_{3/2}$)\,=\,(6.2\,$\pm$\,0.2)\,$\times$\,10$^{12}$ cm$^{-2}$, and $N$($^2\Pi_{1/2}$)\,=\,(8.0\,$\pm$\,0.3)\,$\times$\,10$^{10}$ cm$^{-2}$. The calculated
synthetic spectra are shown in Fig. \ref{C6H} and they fit  the observed lines of the two ladders remarkably well. 
From the derived column densities and the energy difference of 22\,K between the ladders, it is possible to derive
an interladder temperature of 5.1$\pm$0.3\,K (which neglects the contribution of the
$\nu_{11}(\mu^2\Sigma)$  mode). The total column density of C$_6$H has to take into account the contribution
of the two ladders and of the $\nu_{11}(\mu^2\Sigma)$ vibrational state. Hence, $N$(C$_6$H)=(6.4$\pm$0.2)$\times$10$^{12}$ cm$^{-2}$. 
This value of $N$(C$_6$H) is consistent (within 10\%) with that derived previously \citep[see Table A.1 of][]{Cernicharo2022}.
The fraction of C$_6$H molecules in the 
$^2\Pi_{3/2}$, $^2\Pi_{1/2}$, and $\nu_{11}$ states is 96.8\%, 1.3\%, and 1.9\%, respectively. 
Adopting the energy of 15.8\,K
for the $\nu_{11}(\mu^2\Sigma)$ level \citep{Zhao2011}, then the interladder temperature between
the ground $^2\Pi_{3/2}$ state and this vibrational level is 4.0$\pm$0.3\,K.

Our failure to reproduce the intensity of the lines when considering the two ladders simultaneously is due to the fact that
the rotational levels of the $\nu_{11}(\mu^2\Sigma)$ state have to be merged with those of $^2\Pi$ ground state. The partition function of C$_6$H
has to take into account all levels. Taking into account that this vibrational mode is slightly below in energy than 
the $^2\Pi_{1/2}$ ladder
(15.8 versus 22\,K), a significant fraction of the level population in this ladder is transfered to the vibrational mode. 
Consequently, for future analyses of C$_6$H, it will be necessary to consider that the molecule has three ladders: the two
spin components of the ground and the low-lying vibrational mode $\nu_{11}(\mu^2\Sigma)$. 

\subsection{Detections in other sources}

We also detected C$_6$H in the $\nu_{11}(\mu^2\Sigma)$ vibrational state in four 
additional cold dense clouds, namely Lupus-1A, L1527, L1495B, and L483. These sources were 
observed with the Yebes\,40m telescope in the Q band, selected because 
they exhibit strong emission in the lines of cyanopolyynes and radicals, and they serve as a comparative 
benchmark with TMC-1. Details of these observations can be found 
in \cite{Agundez2023}. 
The coordinates of these sources are given in Table \ref{table:lines_clouds}.
A selection of the detected lines of C$_6$H $\nu_{11}(\mu^2\Sigma)$ is shown 
in Fig.\,\ref{fig:lines_clouds}, while the line parameters for all observed transitions
are listed in Table\,\ref{table:lines_clouds}. 
We derived the column densities of C$_6$H in the $\nu_{11}(\mu^2\Sigma)$ state towards
Lupus-1A and L1527 through rotational diagrams, assuming that the source fills the beam of
the telescope. In the case of Lupus-1A and L1527, there are enough lines to derive a rotational temperature, while for L1495B and L483, there is only a low number of lines; therefore, we fixed the rotational temperature to the values determined for C$_6$H in the ground vibrational state \citep{Agundez2023}. In Lupus-1A, we derived $T_{rot}$\,=\,10.5$\pm$2.1\,K and $N$(C$_6$H, $\nu_{11}(\mu^2\Sigma)$)\,=\,(1.4\,$\pm$\,0.5)\,$\times$\,10$^{11}$ cm$^{-2}$; while in L1527, we obtained $T_{rot}$\,=\,30.8$\pm$17.1\,K and $N$(C$_6$H, $\nu_{11}(\mu^2\Sigma)$)\,=\,(1.3\,$\pm$\,0.4)\,$\times$\,10$^{11}$ cm$^{-2}$. For Lupus-1A, the rotational temperature derived is consistent with the value of 10.7\,$\pm$\,0.7 K derived for C$_6$H by \cite{Agundez2023}; while in the case of L1527, the rotational temperature is somewhat higher, although it is consistent within the uncertainty, with the value of 19.6\,$\pm$\,3.4 K derived for C$_6$H by \cite{Agundez2023}. For L1495B and L483, we adopted rotational temperatures of 7.0 K and 8.3 K, respectively, and we derived column densities for C$_6$H $\nu_{11}$($\mu^2\Sigma$) of 6.1\,$\times$\,10$^{10}$ cm$^{-2}$ and 4.5\,$\times$\,10$^{10}$ cm$^{-2}$, respectively.

The column density ratios between the $\nu_{11}(\mu^2\Sigma)$ state and the ground vibrational state 
of C$_6$H are 3.8\,\%, 14.8\,\%, 4.1\,\%, and 6.0\,\% for Lupus-1A, L1527, L1495B, and L483, respectively.
These population fractions correspond to interladder temperatures of 4.8, 8.3, 5.0, and 5.6\,K, respectively.
The vibronic state $\nu_{11}$($\mu^2\Sigma$) is therefore more populated in these four clouds than 
in TMC-1, especially in the case of L1527. This latter cloud is known to be significantly warmer 
than the others \citep{Sakai2008,Agundez2023}. It thus seems that the fraction of C$_6$H in 
the vibronic state is correlated with the gas kinetic temperature and the volume density, which suggests 
that the rotational levels within this vibrationally excited state are mainly populated 
through collisions. That is to say, the warmer and denser the source, the larger 
the fractional population of the vibrationally excited state.
However, the exact way in which these levels are excited is likely the 
result of a complex interplay of collisional and radiative processes involving 
the $^2\Pi_{3/2}$, $^2\Pi_{1/2}$, and $\nu_{11}(\mu^2\Sigma)$ ladders.

\section{Discussion and conclusions}\label{sec:discussion}

We report the first detection of an excited vibrational state in cold starless cores. This opens up the possibility that other low-lying vibrationally excited states
of abundant species are detected in these types of sources. We also searched for the lines of the $^2\Delta$ vibronic state, 
observed towards IRC+10216 \citep{Cernicharo2008}. However, as expected from its
energy of $\sim$60\,K \citep{Zhao2011}, these lines have not been detected in TMC-1 or in any of the other sources. 
IRC+10216 represents a warmer environment and 
infrared radiative pumping plays
an important role in the population of the vibrational states of abundant molecules 
\citep{Cernicharo2008,Cernicharo2013,Cernicharo2014,Pardo2018}.
Thermal emission from vibrationally excited states of HC$_3$N is often found in warm and hot molecular
clouds \citep{deVicente2000,Peng2017,Taniguchi2022}. However, its lowest bending mode 
is at $\sim$222 cm$^{-1}$
\citep{Bizzocchi2017}. Hence, we do not expect emission from this vibrational level 
in cold sources. 

The detection of lines from the $\nu_{11}(\mu^2\Sigma)$ vibrational mode of C$_6$H in cold dark clouds
indicates that the low energy vibrational levels of abundant molecules have to be considered in the
analysis of sensitive line surveys. Lines arising 
from low-lying vibrationally excited bending states of C$_n$H and C$_n$N radicals are potential candidates.
Some of these species have a low-lying electronic state close to their ground state that could
induce important effects in the energy of their bending vibrational modes 
(see, e.g. Botschwina et al. 1997).
Other molecules have such Renner-Teller interactions lowering
the energies of their bending modes. Among the ones that have been fully characterised in the laboratory, we have
the $\nu_4$ mode of C$_3$H \citep{Yamamoto1990} 
and the $\nu_7$ mode of C$_4$H \citep{Yamamoto1987,Guelin1987b,Cooksy2015}. The latter has been observed
in IRC+10216 \citep{Guelin1987b,Cernicharo2000}. However, the rotational transitions arising from these vibrational modes
involve energy levels above 30\,K and, thus, they were not detected in the sources observed in this work.

We have found that the available collisional rates for C$_6$H \citep{Walker2018} require very high volumne 
densities ($n$(H$_2$)$\ge$10$^6$ cm$^{-3}$) to reproduce the observed intensities and the
rotational temperatures derived in our analysis. Refined quantum chemical calculations of these rates
are needed, including the $\nu_{11}$ state, in order to obtain reasonable values for
the volume density from the observation of the C$_6$H radical in interstellar clouds.

\begin{acknowledgements}

We thank Ministerio de Ciencia e Innovaci\'on of Spain (MICIU) for funding support through projects
PID2019-106110GB-I00, PID2019-107115GB-C21 / AEI / 10.13039/501100011033, and
PID2019-106235GB-I00. We also thank ERC for funding
through grant ERC-2013-Syg-610256-NANOCOSMOS. We would like to thank our referee, Prof. S. Yamamoto, for usefull comments
and suggestions.

\end{acknowledgements}

\normalsize
\begin{appendix}
\onecolumn
\section{Improved rotational constants for C$_6$H $\nu_{11}$($^2\Sigma$)}\label{rota_v11}
The $\nu_{11}(\mu^2\Sigma)$ state
is highly perturbed and high distortion constants ($D, H, L$) are needed to reproduce the laboratory frequencies \citep{Gottlieb2010}.
Although the predicted frequencies for the lines of this state within the QUIJOTE line survey fit
the observed ones reasonably well, we have included them in a merged fit to all laboratory and space lines. 
The distortion constants $D, H, L,$ and $M$ for the rotation, and $D$ and $H$ for the
fine structure have been included. The results are given in Table \ref{rotationalconstants} and
correspond to the recommended constants to predict the frequencies of the $\nu_{11}$($^2\Sigma$) state. 
The table also gives the constants derived by \citet{Gottlieb2010} and those derived from the same set of 
measurements but including the distortion constant, $M$.

\begin{table*}
\centering
\caption{Rotational and distortion constants C$_6$H $\nu_{11}(^2\Sigma)$.}
\label{rotationalconstants}
\centering
\begin{tabular}{{cccc}}
\hline
Constant$^a$        & Laboratory$^b$ &Laboratory$^c$ & Laboratory+TMC-1\,$^d$ \\
 (MHz)              &                &               &                        \\
\hline                               
$B$                 & 1394.64000(4)  &1394.63997(6)  &   1394.64004(5)      \\
10$^6$\,$D$         & 46.48(3)       & 46.208(63)    &   46.243(58)           \\
10$^{12}$\,$H$      & 27(7)          &-116(26)       &  -108(25)              \\
10$^{16}$\,$L$      & -37(5)         & 220(44)       &   211(43)              \\
10$^{18}$\,$M$      &                & -1.51(25)     &  -1.47(25)             \\
$\gamma$            & -18.826(2)     & -18.85682(26) & -18.85434(22)          \\
10$^3$\,$\gamma_D$  &   2.927(3)     &  2.9346(31)   &   2.9350(30)           \\ 
10$^6$\,$\gamma_H$  &  -0.1164(9)    & -0.11745(93)  &  -0.11764(89)           \\
10$^{12}$\,$\gamma_L$&   3.67(7)      &  3.730(72)    &   3.747(69)             \\
$N_{lines}$         &   59           & 59            &      71                \\
$\sigma$ (kHz)      &                & 15.6          & 14.1                   \\ 
\hline
\end{tabular}
\tablefoot{\\
\tablefoottext{a}{All constants in MHz.
The uncertainties (in parentheses) are in units of the last 
significant digits.}\\
\tablefoottext{b}{Rotational and distortion constants given by \citet{Gottlieb2010}.}\\
\tablefoottext{c}{Rotational and distortion constants derived from a fit to  the
laboratory data including the distortion constant $M$. The hyperfine constants $b_F$ and $c$ have been
fixed to the values given by \citet{Gottlieb2010}.}\\
\tablefoottext{d}{Rotational and distortion constants derived from a fit to  the
laboratory and TMC-1 data. The hyperfine constants $b_F$ and $c$ have been
fixed to the values given by \citet{Gottlieb2010}. The distortion constant $M$ has
been included to improve the quality of fit.}\\
}
\end{table*}

\section{Line parameters}
Line parameters for all observed transitions were derived by fitting a Gaussian line profile to them
using the GILDAS package. A
velocity range of $\pm$20\,\kms\, around each feature was considered for the fit after a polynomial 
baseline was removed. Negative features produced in the folding of the frequency switching data were blanked
before baseline removal. The derived line parameters for TMC-1 are
given in Table \ref{line_parameters} and in Table\,\ref{table:lines_clouds} for the other sources.

\begin{table*}[h]
\centering
\caption{Observed line parameters for C$_6$H in the $\nu_{11}(\mu^2\Sigma)$ excited and ground states in TMC-1.}
\label{line_parameters}
\begin{tabular}{lcccccccc}
\hline
State           & Transition$^\$$      & & $\nu_{rest}$~$^a$ & $\int T_A^* dv$~$^b$ & v$_{LSR}$       & $\Delta v$~$^c$ & $T_A^*$~$^d$ \\
                &                     & & (MHz)              & (mK\,km\,s$^{-1}$)  & (km\,s$^{-1}$)  & (km\,s$^{-1}$)  & (mK) \\
\hline
$\nu_{11} (^2\Sigma)$& 12-11      &$a$& 33462.231$\pm$0.010 &  3.92$\pm$0.11& 5.83 & 0.77$\pm$0.03&  4.78$\pm$0.07&         \\
                     & 12-11      &$b$& 33479.823$\pm$0.010 &  2.77$\pm$0.07& 5.83 & 0.77$\pm$0.02&  3.40$\pm$0.07&         \\
                     & 13-12      &$a$& 36251.528$\pm$0.010 &  3.78$\pm$0.10& 5.83 & 0.82$\pm$0.03&  4.20$\pm$0.12& A\\
                     & 13-12      &$b$& 36268.909$\pm$0.010 &  2.41$\pm$0.05& 5.83 & 0.69$\pm$0.02&  3.27$\pm$0.07&         \\
                     & 14-13      &$a$& 39040.824$\pm$0.010 &  2.89$\pm$0.07& 5.83 & 0.66$\pm$0.03&  4.11$\pm$0.08&         \\
                     & 14-13      &$b$& 39057.965$\pm$0.010 &  1.95$\pm$0.06& 5.83 & 0.63$\pm$0.02&  2.92$\pm$0.09&         \\
                     & 15-14      &$a$& 41830.108$\pm$0.010 &  2.65$\pm$0.10& 5.83 & 0.67$\pm$0.03&  3.71$\pm$0.11&         \\
                     & 15-14      &$b$& 41847.005$\pm$0.010 &  1.66$\pm$0.09& 5.83 & 0.60$\pm$0.04&  2.61$\pm$0.12&         \\
                     & 16-15      &$a$& 44619.383$\pm$0.010 &  2.30$\pm$0.10& 5.83 & 0.66$\pm$0.03&  3.27$\pm$0.12&         \\
                     & 16-15      &$b$& 44636.019$\pm$0.010 &  1.37$\pm$0.13& 5.83 & 0.68$\pm$0.08&  1.88$\pm$0.12&         \\
                     & 17-16      &$a$& 47408.647$\pm$0.010 &  1.12$\pm$0.12& 5.83 & 0.55$\pm$0.07&  1.92$\pm$0.13&         \\
                     & 17-16      &$b$& 47425.006$\pm$0.010 &  1.03$\pm$0.13& 5.83 & 0.61$\pm$0.09&  1.58$\pm$0.16&         \\
                     &            & &                     &               &              &              &               &         \\
$^2\Pi_{3/2}$        & 23/2-21/2  &$e$& 31881.860$\pm$0.002 &173.68$\pm$0.06& 5.74$\pm$0.01& 1.12$\pm$0.01&146.04$\pm$0.08&         \\
                     &            &$f$& 31885.541$\pm$0.002 &175.02$\pm$0.13& 5.73$\pm$0.01& 1.09$\pm$0.01&150.28$\pm$0.09&         \\
                     & 25/2-23/2  &$e$& 34654.037$\pm$0.002 &156.89$\pm$0.14& 5.74$\pm$0.01& 1.05$\pm$0.01&140.59$\pm$0.07&         \\
                     &            &$f$& 34658.383$\pm$0.002 &156.17$\pm$0.09& 5.74$\pm$0.01& 1.01$\pm$0.01&144.69$\pm$0.09&         \\
                     & 27/2-25/2  &$e$& 37426.192$\pm$0.002 &139.15$\pm$0.05& 5.75$\pm$0.01& 0.97$\pm$0.01&134.46$\pm$0.09&         \\
                     &            &$f$& 37431.255$\pm$0.002 &138.02$\pm$0.14& 5.74$\pm$0.01& 0.94$\pm$0.01&138.84$\pm$0.09&         \\
                     & 29/2-27/2  &$e$& 40198.323$\pm$0.002 &118.21$\pm$0.12& 5.76$\pm$0.01& 0.90$\pm$0.01&123.29$\pm$0.09&         \\
                     &            &$f$& 40204.157$\pm$0.002 &118.74$\pm$0.12& 5.75$\pm$0.01& 0.88$\pm$0.01&127.34$\pm$0.09&         \\
                     & 31/2-29/2  &$e$& 42970.432$\pm$0.002 & 91.80$\pm$0.10& 5.76$\pm$0.01& 0.85$\pm$0.01&100.94$\pm$0.12&         \\
                     &            &$f$& 42977.089$\pm$0.002 & 91.76$\pm$0.10& 5.75$\pm$0.01& 0.82$\pm$0.01&104.66$\pm$0.11&         \\
                     & 33/2-31/2  &$e$& 45742.519$\pm$0.002 & 70.27$\pm$0.12& 5.77$\pm$0.01& 0.80$\pm$0.01& 82.28$\pm$0.14&         \\
                     &            &$f$& 45750.052$\pm$0.002 & 70.28$\pm$0.13& 5.76$\pm$0.01& 0.77$\pm$0.01& 85.29$\pm$0.16&         \\
                                         & 35/2-33/2  &$e$& 48514.584$\pm$0.002 & 51.51$\pm$0.16& 5.77$\pm$0.01& 0.78$\pm$0.01& 62.27$\pm$0.18&         \\
                     &            &$f$& 48523.044$\pm$0.002 & 51.00$\pm$0.15& 5.76$\pm$0.01& 0.75$\pm$0.01& 63.69$\pm$0.17&         \\
                     &            & &                     &               &              &              &               &         \\                                                                                                                                                                                                                                                                
$^2\Pi_{1/2}$        & 23/2-21/2  &$e$& 32095.245$\pm$0.003 &  2.07$\pm$0.09& 5.76$\pm$0.02& 0.75$\pm$0.04&  2.58$\pm$0.08&         \\
                     &            &$f$& 32125.561$\pm$0.003 &  1.42$\pm$0.08& 5.83$\pm$0.01& 0.65$\pm$0.04&  2.07$\pm$0.10&         \\
                     & 25/2-23/2  &$e$& 34887.115$\pm$0.003 &  2.25$\pm$0.07& 5.84$\pm$0.02& 0.98$\pm$0.03&  2.17$\pm$0.08& B       \\
                     &            &$f$& 34917.885$\pm$0.003 &  1.86$\pm$0.06& 5.83$\pm$0.01& 0.79$\pm$0.03&  2.21$\pm$0.08&         \\
                     & 27/2-25/2  &$e$& 37678.938$\pm$0.003 &  1.91$\pm$0.13& 5.80$\pm$0.03& 0.72$\pm$0.06&  2.50$\pm$0.17& C       \\
                     &            &$f$& 37710.199$\pm$0.002 &  1.62$\pm$0.06& 5.89$\pm$0.02& 0.73$\pm$0.03&  2.11$\pm$0.09&         \\
                     & 29/2-27/2  &$e$& 40470.713$\pm$0.003 &  1.61$\pm$0.08& 5.79$\pm$0.02& 0.70$\pm$0.04&  2.17$\pm$0.08&         \\
                     &            &$f$& 40502.502$\pm$0.003 &  1.75$\pm$0.13& 5.95$\pm$0.03& 0.88$\pm$0.08&  1.87$\pm$0.12&         \\
                     & 31/2-29/2  &$e$& 43262.437$\pm$0.003 &  1.36$\pm$0.10& 5.79$\pm$0.03& 0.77$\pm$0.07&  1.65$\pm$0.11&         \\
                     &            &$f$& 43294.790$\pm$0.003 &  1.21$\pm$0.10& 5.85$\pm$0.03& 0.66$\pm$0.07&  1.73$\pm$0.10&         \\
                     & 33/2-31/2  &$e$& 46054.109$\pm$0.003 &  0.80$\pm$0.10& 5.76$\pm$0.03& 0.53$\pm$0.07&  1.42$\pm$0.11&         \\
                     &            &$f$& 46087.062$\pm$0.003 &  0.64$\pm$0.10& 6.01$\pm$0.05& 0.74$\pm$0.12&  0.82$\pm$0.13&         \\
                                         & 35/2-33/2  &$e$& 48845.727$\pm$0.003 &  0.53$\pm$0.10& 5.81$\pm$0.05& 0.48$\pm$0.09&  1.03$\pm$0.18&         \\
                     &            &$f$& 48879.315$\pm$0.003 &  0.86$\pm$0.12& 5.89$\pm$0.05& 0.75$\pm$0.12&  1.08$\pm$0.17&         \\
\hline
\end{tabular}
\tablefoot{
\tablefoottext{\$}{The $a$ and $b$ labels refers to the $J=N+1/2$ and
$J=N-1/2$ fine components of the $\nu_{11}(^2\Sigma)$ state.
The $e$ and $f$ labels correspond to the two lambda-doubling components
of the $^2\Pi_{3/2}$ and $^2\Pi_{1/2}$ ladders of the $^2\Pi$ ground
electronic state.}
\tablefoottext{a}{Predicted frequencies for the transitions of C$_6$H in MHz. These
frequencies are the observed ones for the $\nu_{11}$ vibrational state assuming a v$_{LSR}$ of 5.83
km\,s$^{-1}$ (see text).}
\tablefoottext{b}{Integrated line intensity in mK\,km\,s$^{-1}$.} 
\tablefoottext{c}{Line width at half intensity using a Gaussian fit in the line profile (in $km~s^{-1}$).
For the $^2\Pi_{3/2}$ and $^2\Pi_{1/2}$ ladders of the ground state the lines are slightly broadened
by the presence of two hyperfine components, which in addition produce a 
shift of a few kHz to the center frequency. The broadening decreases when $J$ increases.}
\tablefoottext{d}{Antenna temperature (in $mK$).}
\tablefoottext{A}{Only data from the observations with a frequency switching of 8 MHz.}
\tablefoottext{B}{The line is partially affected by a negative feature.}
\tablefoottext{C}{Only data from the observations with a frequency switching of 10 MHz.}
}\\
\end{table*}

\begin{table*}[h]
\centering
\caption{Observed line parameters for C$_6$H $\nu_{11}(\mu^2\Sigma)$ in Lupus-1A, L1527, L1495B, and L483.}
\label{table:lines_clouds}
\begin{tabular}{cccccc}
\hline
Transition      & $\nu_{calc}$ & $\int T_A^* dv$       & v$_{LSR}$       & $\Delta v$        & $T_A^*$ \\
                      & (MHz)           & (mK\,km\,s$^{-1}$) & (km\,s$^{-1}$)  & (km\,s$^{-1}$)  & (mK) \\
\hline
\\
\multicolumn{6}{c}{Lupus-1A ($\alpha_{2000}$=15$^h$42$^m$52.4$^s$, $\delta_{2000}$=-34$^o$ 07' 53.5")} \\
\hline
12-11 $a$ & 33462.222 &  6.94$\pm$0.32 & 5.07$\pm$0.02 & 0.64$\pm$0.03 & 10.20$\pm$0.46 \\
12-11 $b$ & 33479.823 &  5.13$\pm$0.34 & 5.12$\pm$0.02 & 0.67$\pm$0.05 &  7.18$\pm$0.48 \\
13-12 $a$ & 36251.521 &  6.09$\pm$0.47 & 5.03$\pm$0.02 & 0.58$\pm$0.05 &  9.95$\pm$0.60 \\
13-12 $b$ & 36268.907 &  4.82$\pm$0.50 & 5.05$\pm$0.04 & 0.74$\pm$0.10 &  6.15$\pm$0.56 \\
14-13 $a$ & 39040.815 &  5.97$\pm$0.39 & 5.04$\pm$0.02 & 0.53$\pm$0.03 & 10.62$\pm$0.68 \\
14-13 $b$ & 39057.968 &  5.34$\pm$0.50 & 5.05$\pm$0.03 & 0.56$\pm$0.07 &  8.89$\pm$0.81 \\
15-14 $a$ & 41830.101 &  4.50$\pm$0.52 & 5.05$\pm$0.03 & 0.49$\pm$0.07 &  8.62$\pm$0.92 \\
15-14 $b$ & 41847.006 &  2.81$\pm$0.46 & 5.11$\pm$0.04 & 0.37$\pm$0.13 &  7.09$\pm$0.89 \\
16-15 $a$ & 44619.378 &  5.37$\pm$0.59 & 5.06$\pm$0.03 & 0.53$\pm$0.07 &  9.49$\pm$1.10 \\
16-15 $b$ & 44636.019 &  3.80$\pm$0.65 & 5.16$\pm$0.06 & 0.66$\pm$0.14 &  5.38$\pm$1.03 \\
17-16 $a$ & 47408.645 &  3.80$\pm$0.82 & 5.11$\pm$0.05 & 0.50$\pm$0.13 &  7.13$\pm$1.85 \\
17-16 $b$ & 47425.006 &  4.07$\pm$1.40 & 5.20$\pm$0.10 & 0.60$\pm$0.25 &  6.32$\pm$2.43 \\
\\
\multicolumn{6}{c}{L1527 ($\alpha_{2000}$=4$^h$39$^m$53.9$^s$, $\delta_{2000}$=+26$^o$ 03' 11.0")} \\
\hline
12-11 $a$ & 33462.222 &  3.88$\pm$0.41 & 5.83$\pm$0.04 & 0.79$\pm$0.10 &  4.59$\pm$0.52 \\
12-11 $b$ & 33479.823 &  2.74$\pm$0.31 & 5.97$\pm$0.04 & 0.69$\pm$0.08 &  3.72$\pm$0.44 \\
13-12 $a$ & 36251.521 &  6.48$\pm$0.47 & 5.88$\pm$0.03 & 0.91$\pm$0.08 &  6.71$\pm$0.59 \\
13-12 $b$ & 36268.907 &  4.19$\pm$0.37 & 5.85$\pm$0.03 & 0.69$\pm$0.07 &  5.71$\pm$0.54 \\
14-13 $a$ & 39040.815 &  6.26$\pm$0.35 & 5.88$\pm$0.02 & 0.80$\pm$0.05 &  7.33$\pm$0.47 \\
14-13 $b$ & 39057.968 &  5.48$\pm$0.47 & 5.84$\pm$0.04 & 1.05$\pm$0.11 &  4.92$\pm$0.55 \\
15-14 $a$ & 41830.101 &  4.59$\pm$0.53 & 5.84$\pm$0.03 & 0.52$\pm$0.08 &  8.23$\pm$0.92 \\
15-14 $b$ & 41847.006 &  5.09$\pm$0.54 & 5.97$\pm$0.03 & 0.64$\pm$0.08 &  7.51$\pm$0.83 \\
16-15 $a$ & 44619.378 &  5.65$\pm$0.39 & 5.79$\pm$0.02 & 0.62$\pm$0.05 &  8.62$\pm$0.83 \\
16-15 $b$ & 44636.019 &  4.10$\pm$0.58 & 5.89$\pm$0.05 & 0.74$\pm$0.13 &  5.24$\pm$0.84 \\
17-16 $a$ & 47408.645 &  4.85$\pm$0.58 & 5.88$\pm$0.04 & 0.64$\pm$0.08 &  7.15$\pm$1.02 \\
17-16 $b$ & 47425.006 &  3.98$\pm$0.76 & 5.90$\pm$0.06 & 0.69$\pm$0.17 &  5.40$\pm$1.17 \\
\\
\multicolumn{6}{c}{L1495B ($\alpha_{2000}$=4$^h$15$^m$41.8$^s$, $\delta_{2000}$=+28$^o$ 47' 46.0")} \\
\hline
12-11 $a$ & 33462.222 &  1.20$\pm$0.37 & 7.44$\pm$0.13 & 0.70$\pm$0.23 &  1.60$\pm$0.51 \\
12-11 $b$ & 33479.823 &  1.42$\pm$0.36 & 7.87$\pm$0.10 & 0.78$\pm$0.22 &  1.72$\pm$0.48 \\
13-12 $a$ & 36251.521 &  2.02$\pm$0.38 & 7.58$\pm$0.07 & 0.78$\pm$0.18 &  2.43$\pm$0.41 \\
13-12 $b$ & 36268.907 &  1.64$\pm$0.39 & 7.69$\pm$0.08 & 0.70$\pm$0.22 &  2.19$\pm$0.51 \\
14-13 $a$ & 39040.815 &  1.20$\pm$0.29 & 7.67$\pm$0.06 & 0.47$\pm$0.11 &  2.40$\pm$0.53 \\
14-13 $b$ & 39057.968 &  1.46$\pm$0.42 & 7.76$\pm$0.09 & 0.57$\pm$0.16 &  2.40$\pm$0.73 \\
15-14 $a$ & 41830.101 &  2.13$\pm$0.49 & 7.64$\pm$0.04 & 0.43$\pm$0.16 &  4.68$\pm$0.91 \\
\\
\multicolumn{6}{c}{L483 ($\alpha_{2000}$=18$^h$17$^m$29.8$^s$, $\delta_{2000}$=-4$^o$ 39' 38.3")} \\
\hline
12-11 $a$ & 33462.222 &  1.59$\pm$0.23 & 5.16$\pm$0.07 & 0.89$\pm$0.13 &  1.68$\pm$0.30 \\
13-12 $b$ & 36268.907 &  1.19$\pm$0.18 & 5.35$\pm$0.06 & 0.70$\pm$0.12 &  1.61$\pm$0.24 \\
14-13 $a$ & 39040.815 &  1.53$\pm$0.26 & 5.29$\pm$0.07 & 0.83$\pm$0.15 &  1.73$\pm$0.36 \\
15-14 $b$ & 41847.006 &  0.90$\pm$0.19 & 5.26$\pm$0.05 & 0.47$\pm$0.09 &  1.82$\pm$0.35 \\
16-15 $b$ & 44636.019 &  1.62$\pm$0.34 & 5.43$\pm$0.07 & 0.63$\pm$0.14 &  2.43$\pm$0.58 \\
\hline
\end{tabular}
\end{table*}

\end{appendix}

\begin{thebibliography}{} 
\tiny
\bibitem[Ag\'undez et al.(2023)]{Agundez2023} Ag\'undez, M., Marcelino, N., Tercero, B., et al. 2023, \aap, 677, A106
\bibitem[Bizzocchi et al.(2017)]{Bizzocchi2017}Bizzocchi, L., Tamassia, F., Lass, J., et al. 2017, \apj, 233, 11 
\bibitem[Botschwina et al. (1997)]{Botschwina1997}Botschwina, P., Horn, M., Markey, K. \& Oswald, R. 1997, Mol. Phys., 92, 381
\bibitem[Brown et al.(1999)]{Brown1999}Brown, S.T., Rienstra-Kiracofe, R. \& Schaefer , H.F. III 1999, J. Phys. Chem. A, 103, 4065
\bibitem[Cao \& Peyerimhoff (2001)]{Cao2001}Cao, Z. \& Peyerimhoff, S.D. 2001, Phys. Chem. Chem. Phys., 3, 1403
\bibitem[Cernicharo(1985)]{Cernicharo1985} Cernicharo, J. 1985, Internal IRAM report (Granada: IRAM)
\bibitem[Cernicharo et al.(1987)]{Cernicharo1987a} Cernicharo, J. Gu\'elin, M., Menten, K. \& Walmsley, C.M. 1987, \aap, 181, L1
\bibitem[Cernicharo \& Gu\'elin(1987)]{Cernicharo1987b} Cernicharo, J. \& Gu\'elin, M. 1987, \aap, 176, 299
\bibitem[Cernicharo et al.(2000)]{Cernicharo2000} Cernicharo, J., Gu{\'e}lin, M. \& Kahane, C. 2000, \aap\,S.S., 142, 181 
\bibitem[Cernicharo et al.(2008)]{Cernicharo2008} Cernicharo, J., Gu{\'e}lin, M., Ag{\'u}ndez, M., et al.\ 2008, \apjl, 688, L83
\bibitem[Cernicharo(2012)]{Cernicharo2012} Cernicharo, J., 2012, in ECLA 2011: Proc. of the European Conference on Laboratory Astrophysics,
EAS Publications Series, 2012, Ed.: C. Stehl, C. Joblin, \& L. d'Hendecourt (Cambridge: Cambridge Univ. Press),
251; \texttt{https://nanocosmos.iff.csic.es/?page$\_$id=1619}
\bibitem[Cernicharo et al.(2013)]{Cernicharo2013}Cernicharo, J., Daniel, F., Castro-Carrizo, A. et al. 2013, \apj, 778, L25
\bibitem[Cernicharo et al.(2014)]{Cernicharo2014}Cernicharo, J., Teyssier, D., Quintana-Lacaci, G. et al. 2014, \apj, 796, L21
\bibitem[Cernicharo et al.(2021a)]{Cernicharo2021a} Cernicharo, J., Ag\'undez, M., Kaiser, R., et al. 2021a, \aap, 652, L9 
\bibitem[Cernicharo et al.(2021b)]{Cernicharo2021b} Cernicharo, J., Cabezas, C., Endo, Y., et al. 2021b, \aap, 646, L3 
\bibitem[Cernicharo et al.(2022)]{Cernicharo2022}Cernicharo, J., Fuentetaja, R., Ag\'undez, M. et al. 2022, \aap, 663, L9 
\bibitem[Cooksy et al. (2015)]{Cooksy2015}Cooksy, A.~L., Gottlieb, C.~A., Killian, T.~C. et al. 2015, \apjs, 216, 30
\bibitem[de Vicente et al.(2000)]{deVicente2000}de Vicente, P., Mart\'in-Pintado, J. \& Colom, P. 2000, \aap, 361, 1058
\bibitem[Foss\'e et al.(2001)]{Fosse2001} Foss\'e, D., Cernicharo, J., Gerin, M., Cox, P. 2001, \apj, 552, 168
\bibitem[Fuentetaja et al.(2023)]{Fuentetaja2023} Fuentetaja, R. et al. 2023, in preparation
\bibitem[Gottlieb et al.(2010)]{Gottlieb2010} Gottlieb, C.~A., McCarthy, M.~C., \& Thaddeus, P.\ 2010, \apjs, 189, 261. 
\bibitem[Gu\'elin et al.(1987)]{Guelin1987}Gu\'elin, M., Cernicharo, J., Kahane, C. et al. 1987, \aap, 175, L5 
\bibitem[Gu\'elin et al.(1987b)]{Guelin1987b}Gu\'elin, M., Cernicharo, J., Navarro, S. et al. 1987b, \aap, 182, L37
\bibitem[Kaifu et al.(2004)]{Kaifu2004} Kaifu, N., Ohishi, M., Kawaguchi, K., et al. 2004, PASJ, 56, 69
\bibitem[Linnartz et al.(1999)]{Linnartz1999} Linnartz, H., Motylewski, T., Vaizert, O., et al.\ 1999, Journal of Molecular Spectroscopy, 197, 1
\bibitem[McCarthy et al.(2006)]{McCarthy2006} McCarthy, M.~C., Gottlieb, C.~A., Gupta, H., et al.\ 2006, \apjl, 652, L141
\bibitem[M\"uller et al.(2005)]{Muller2005} M\"uller, H.~S.~P., Schl\"oder, F., Stutzki, J., Winnewisser, G. 2005, \jmst, 742, 215 
\bibitem[Murakami et al.(1988)]{Murakami1988}Murakami, A., Kawaguchi, K. \& Saito, S. 1988, Pub. Astron. Soc. Japan, 39, 189
\bibitem[Pardo et al.(2001)]{Pardo2001} Pardo, J.~R., Cernicharo, J., Serabyn, E. 2001, IEEE Trans. Antennas and Propagation, 49, 12
\bibitem[Pardo et al.(2018)]{Pardo2018} Pardo, J.~R., Cernicharo, J., Velilla-Prieto, L. et al. 2018, \aap, 615, L4
\bibitem[Pearson et al.(1988)]{Pearson1988}Pearson, J.~C., Gottlieb, C.~A., Woodward, D.~R. \& Thaddeus, P. 1988, \aap, 189, L3
\bibitem[Peng et al.(2017)]{Peng2017}Peng, Y., Qin, S.L., Schilke, P. et al. 2017, \apj, 837, 49
\bibitem[Sakai et al. (2008)]{Sakai2008}Sakai, N., Sakai, T., Hirota, T. \& Yamamoto, S. 2008, \apj, 672, 371
\bibitem[Suzuki et al.(1986)]{Suzuki1986} Suzuki, H., Ohishi, M., Kaifu, N., et al.\ 1986, \pasj, 38, 911
\bibitem[Taniguchi et al.(2022)]{Taniguchi2022}Taniguchi, K., Tanak, K.E.I., Zhang, Y. et al. 2022, \apj, 931, 99
\bibitem[Tercero et al.(2021)]{Tercero2021} Tercero, F., L\'opez-P\'erez, J. A., Gallego, et al. 2021, \aap, 645, A37
\bibitem[Walker et al.(2018)]{Walker2018} Walker, K.~M., Lique, F., \& Dawes, R.\ 2018, \mnras, 473, 1407 
\bibitem[Woon(1995)]{Woon1995} Woon, D.~E.\ 1995, Chemical Physics Letters, 244, 45. 
\bibitem[Yamamoto et al.(1987)]{Yamamoto1987}Yamamoto, S. Shuji, S., Gu\'elin, M. et al. 1987, \apj, 323, L149
\bibitem[Yamamoto et al.(1990)]{Yamamoto1990}Yamamoto, S. Shuji, S., Suzuki, H. et al. 1990, \apj, 348, 363
\bibitem[Zhao et al.(2011)]{Zhao2011} Zhao, D., Haddad, M.~A., Linnartz, H., et al.\ 2011, \jcp, 135, 044307. 
\end{thebibliography}
\end{document}